\documentstyle[12pt]{article}
\begin{document}

\setlength{\baselineskip}{2em} %\setlength{\baselineskip}{1.0cm}
\begin{titlepage}
\begin{center}
\null\vskip-1truecm

\vskip2.2truecm
{\large\bf $\lambda$ -TYPE PHASE TRANSITION FOR A WEAKLY INTERACTING
BOSE GAS}\\

\vskip2.2truecm
{\bf I.
Cabrera}$^{1}$,
{\bf H. Perez Rojas}$^{1,2}$

\vskip 1.0 cm
$^1${\it  Grupo de Fisica
Te\'orica, ICIMAF,
Calle E No. 309, Vedado, La Habana 4,
Cuba}\\

$^2${\it International Centre for Theoretical Physics,\\
P.O. Box 586 34100 Trieste, Italy}
\bigskip

\begin{abstract}
A mechanism generating a $\lambda$-type behavior in the specific 
heat of a Bose gas near the critical temperature $T_c$ is discussed.
It is shown to work for a general class of quasiparticle spectra, and 
especially, for the
Bogoliubov's model of a weakly interacting Bose
gas, where a
temperature-dependent quasi-particle
spectrum is obtained.

\end{abstract}
\end{center}
\end{titlepage}
\newpage
\section{Introduction}

The first experimental measurement of
the low temperature specific heat of He$^4$ was made by Keesom in 1927-8
\cite{Keesom}
who
became  impressed by its $\lambda$ shape and by its singular behavior
at the critical temperature where superfluid behavior started
to manifest.
In his book on Statistical Mechanics, \cite{Feynman} Feynman refers to the
$\lambda$
point behavior as one of the unsolved problems of superfluidity theory,
and expresses his view that {\it perhaps part of the explanation of the
lambda transition
involves Bose condensation}
since the $c_{v}(T)$ -curve for an
ideal
Bose gas shows a peaked form at the critical temperature,
although it has not a divergent behavior. We are not aware about any
satisfactory model explaining this behavior, except the one presented in
the paper by D.M.Ceperley \cite{Cep} in which it is simulated the boson
 system by
means of the Monte Carlo techniques
using
path integrals. In that paper, a good
agreement is obtained among the simulation and the experimental
measurements of the $He^4$ specific heat.

In our present letter we want to discuss a mechanism for generating 
a $\lambda$-type behavior in the specific heat of a Bose gas. We consider at
first some models for the phase transition of symmetry restoration 
in temperature relativistic systems
with spontaneous symmetry breaking, which exhibit a $\lambda$-type 
behavior for $c_v$, and then we study the similar problem 
in the non-relativistic case. We
show that for the {\it weakly interacting
Bose gas} one can find such a divergent behavior of the specific heat as
being produced by the condensate. Obviously, such a system is far from 
being a satisfactory model for the
superfluid, in which the interactions cannot be considered weak at
all. However, as the mechanism is essentially an infrared (long wavelength) 
effect, it may provide insights  in understanding what 
happen in a more exact
model of  superfluidity. Also, the behavior of the specific heat for
the weakly interacting Bose gas may become
interesting in connection with the recent experimentally observed Bose
condensate, although such problem would require to consider the particles
in an
external field, provided by the magnetic trap.

To start with  we consider two  finite-temperature quantum field
models with spontaneous symmetry breaking (SSB). For the latter, one can
find that a singular behavior of the specific heat occurs at the
symmetry restoration temperature, due to the temperature-dependent mass,
proportional to the symmetry-breaking parameter. We prove that a similar
phenomenon occurs for boson systems whose quasiparticle spectra have some 
specific non-vanishing temperature-dependent spectra in the infrared limit.

We will take either the
simple model of the scalar field or the Abelian Higgs model \cite{Dolan},
\cite{Linde}. One
can write the effective potential as
\begin{equation}
V(\xi) = \frac{\lambda \xi}{4} - \frac{a^2 \xi^2}{2} + V(T, \xi),
\end{equation}
\noindent
where $V(\xi)$ is the sum of tadpole diagrams. The spectrum is
$\epsilon (p)_i = c\sqrt{(p^2 + M^2_i c^2}$, where $i = 1$ for the
scalar and $i= 1,2,3,4$ for the multiplet in the scalar-vector Higgs
model where, $M_1 = \lambda \xi$, and
$M_{2,3,4} = g \xi/2$ and $g$ is the scalar-vector coupling constant. In
the
high temperature limit is $V(T, \xi) = \alpha T^2 \xi^2/2$, the
extremum of $V(\xi)$ leading for $T < T_c$
to a dependence $M(T)_i = \kappa_i \sqrt{T^2_c - T^2}$, $T_c$ being the
critical temperature for symmetry restoration. The infrared contribution
to the thermodynamic potential in the limit $T \gg M(T)_i$ is $\Delta
\Omega \simeq \frac{\sum c^3 M^3_i T}{12 \pi h^3}$. One obtains then a
specific heat
\begin{equation}
\Delta c_v = - T\frac{\partial^2 \Omega}{\partial T^2} \simeq \sum
c^3 \kappa_i^3
\frac{T^4}{2 \pi h^3\sqrt{T^2_c - T^2}},
\end{equation}
\noindent
which diverges for $T \to T_c$, showing a $\lambda$-type behavior.

In general, we may argue that  any quasi-particle 
spectrum 
of the Bose gas whose infrared limit behaves as $\varepsilon ({\bf p})
= a \sqrt{\delta^2 + p^2}$, where $\delta = 
b \sqrt{1 - (T/T_c)^{\gamma}}$,
and  $\gamma > 1$, leads to a $\lambda$-type behavior of $c_v$.
One can obtain that result from the infrared term in the 
calculations made in \cite{Dolan}.
But also, by taking $T$ close enough to $T_c$ to make $\delta$ arbitrary
small, one can choose
some small momentum $\eta \gg \delta$. Then the integral 
over $p$ in the thermodynamic potential $\Omega =\frac V{\beta \hbar ^3}
\int \frac{d^3{\bf p}}{(2\pi )^3}\ln [1-\exp (-\beta \varepsilon 
({\bf p}))]$ can be decomposed
in the sum $\int_0^{\eta} + \int_{\eta}^{\infty}$, from which 
the infrared 
contribution as $\Delta \Omega = \delta^3 T/12 \pi h^3$ is obtained, leading
to the $\lambda$-type behavior for $c_v = 
-T \partial^2 \Omega/\partial T^2$.

As we shall see, for the weakly interacting Bose gas  the previous mechanism
works, if we take $a \sim \delta$
and $\delta^2 \simeq n_0 (T)$, $n_0$ being the Bose-Einstein condensate 
particle density. Obviously,
$T \leq T_c$, where $T_c$ is the critical temperature for condensation and
to describe the system we use a version of the
Bogoliubov's model \cite{Bogoliubov} valid near $T_c$, which
leads to a spectrum having the abovementioned infrared limit.

\section{The Bogoliubov's model near the critical point.}

We shall assume an weakly interacting Bose gas under
the approximate
conditions $a$/$\lambda \ll 1$ and $na^3 \ll 1$
where $a$
 is the scattering length of the two body interaction, $\lambda =
h/\sqrt{2 \pi m T}$ the
thermal wavelength of the particles and $n$ the particle density in the
system  considered. We will start from the quantized field
Hamiltonian expanded in terms of the
quantities $a_{\bf p}^{+}$, $a_{\bf p}^{}$
($\bf p \neq 0)$ as in the usual Bogoliubov's procedure
\cite{Bogoliubov}. In what follows we refer briefly to the
details of this model by following Pathria \cite{Pathria}.

Our case differs from Bogoliubov's one by two facts :1-) as our
temperature interval is
close to the critical temperature, we cannot consider in all cases that
the number of
particles in states with
${\bf p}\neq 0$ is always much more smaller
than $n_o$, number of particles in the
condensate. 2-)The usual approximation $n_o \simeq N$ cannot be made.

Thus we start also from the Hamiltonian of the quantized field for
spinless bosons, we assume momentum conservation in the interactions and
name $U({\bf r})$ the repulsive
potential of the two body interaction,
$U_{\bf p_1,p_2}^{\bf p'_1,p'_2} = \frac{1}{V}\int \int d^3 {\bf
r} e^{\frac{i {\bf p} \cdot {\bf r}}{\hbar}} U ({\bf r})$

The scattering length can be written, then, as $a=\frac{ mV}{4\pi \hbar
^2}U_{\bf p_1,p_2}^{\bf p'_1,p'_2} $, and as we are considering a
repulsive potential always $a > 0$. For ${\bf p}=0$
(zero momentum transfer in the colision) $U_o = \int d^3 {\bf r} U({\bf
r})$.

In the temperature
interval we are considering the momenta
are very small and we can assume that the momentum transfer in
each collision is almost zero ${\bf p}\simeq 0 $. For that reason
it is possible to express approximately the matrix elements by using
$U_o$. Then the
Hamiltonian is,
\begin{equation}
\hat H=\sum \frac{{\bf p}^2}{2m}a_{\bf p}^{+}a_{\bf
p} + \frac{U_o}{2V}\sum a_{{\bf p'}_1}^{+} a_{{\bf p'}_2}^{+}
a_{{\bf p_2}} a_{{\bf p_1}}. \label{segu}
\end{equation}

The usual Bogoliubov's procedure is, by starting from the fact that
the occupation number of the ground state $n_0$ is a macroscopic number,
to do the approximation
$a_o a_o^{+}\simeq a_o a_o \simeq n_o^2$
which allows us to take each one of these creation and
annihilation operators as c-numbers proportional to
$\sqrt{n_o}$. Then the second term in the Hamiltonian can be  expanded, its
 lowest-order term
being $a_o^{+}a_o^{+}a_o a_o=n_o^2-n_o$
and the lowest-order ground state energy is
$E_o=\frac{U_o}{2V}[n_o^2-n_o]$.
The next order terms to be included leads to the term
$\sum\limits_{\bf p \neq 0}\{a_{\bf p}^{+}a_{\bf p}^{+} + a_{\bf p}a_{-\bf
p} +
4a_{\bf p}^{+} a_{-\bf p} \}n_o $ and our final Hamiltonian is

\begin{eqnarray}
\hat H &=&\left. \frac{U_o}{2V}[n_o^2-n_o] + \sum\limits_ {{\bf p}
\neq 0}\frac{{\bf p}^2}{2m} a_{\bf p}^{+} a_{\bf p}\right. \label{i}\\
&&\left. +\frac{U_o n_o}{2V}\sum\limits_{{\bf p}\neq 0}\{a_{\bf p}^{+}
a_{-{\bf p}}^{+} + a_{\bf p} a_{-{\bf p}}+4a_{{\bf p}}^{+}
a_{-{\bf p}}\}\right. . \nonumber
\end{eqnarray}

The usual Bogoliubov's procedure is, by
considering that $n_0 \sim N$, (which ceases to be valid near the
transition temperature) to replace $N$ by $n_0$ in the in the third
term in (\ref{i}) (this means to neglect one term of fourth order in the
quantities  $a_{\bf
p}^+, a_{\bf p}$), and
after
introducing the relation $\sum\limits_{\bf p \neq 0} a_{\bf p}^{+} a_{\bf
p} + n_o = N$ in the first, leads to the cancellation of a term
$\frac{U_0 n_0}{V} a_{\bf p} a^+_{\bf -p}$ in the last bracket of
(\ref{i}).

Thus, our assumptions lead simply to change the coefficient of the last
term in
curly brackets from 2 to 4. The next step is to make the usual Bogoliubov
transformation
$a_{\bf p} = (b_{\bf p} - \alpha_{\bf p} b_{-\bf p}^{+})/\sqrt{1 -
\alpha_{{\bf p}^2}}$ and $a_{\bf p}^{+}=
(b_{\bf p}^{+} - \alpha_{\bf p}
b_{-\bf p})/\sqrt{1-\alpha_{\bf p}^2}$
and as a result of it we get the diagonalized Hamiltonian,

\begin{equation}
\hat H=E_o+\sum \limits_{{\bf p}\neq 0}\varepsilon (
{\bf p})b_{\bf p}^{+} b_{\bf p},
\end{equation}

\noindent where $E_o = \frac{U_o}{2V}[n_o^2-n_o]-\frac{U_o
n_o}{2V}\sum\limits_{\bf p \neq
0}\alpha_{\bf p}$ and $\alpha_{\bf p}=\frac{ V}{
U_on_o}[4\frac{U_on_o}{2V} + \frac{{\bf
p}^2}{2m}-\varepsilon ({\bf p})]$, and if we name $K=\frac{U_o n_0}{2V}$
then

\begin{equation}
\varepsilon ({\bf p})=\sqrt{12K^2 + 8K\frac{{\bf p}^2}{2m
}+(\frac{{\bf p}^2}{2m})^2} , \label{miesp}
\end{equation}

\noindent
is the spectrum of the new Bose quasiparticles representing the
elementary excitations of the system, where
$b_{\bf
p}^{+},
b_{\bf p}$ are their creation and
annihilation operators.

In Bogoliubov's spectrum
the term
$12 K^2$ is not
present. In our model the long wavelength limit is obviously not linear in
$p$. As
$K \ll k T$ is  very small (for $n_0 \simeq 10^{16}, K \simeq 10^{-12}$ eV) 
it can be usually
neglected,
but it is able to produce
the typical $\lambda$-type
divergent behavior of the specific heat near $T_c.$
As $\lim_{{\bf p}\to 0}\varepsilon ({\bf p}
)= 2\sqrt{3}K $, the parameter $K$ formally behaves as the analog of a
rest energy in
relativistic dynamics. This
''rest energy'' has the remarkable property that it decreases with 
temperature and
goes to zero for $T \to T_c$. Actually, $K$ is proportional to the symmetry
breaking
parameter $n_0$, which is the condensate, and condensation means some
(first order) gauge symmetry breaking \cite{Fannes}.
Thus $K$ arises as some
sort
of non-relativistic analog of the temperature scalar and  abelian
Higgs models mentioned above. An explicit computation of $c_v$ for
this case shows a divergent behavior close to the critical temperature, as
it is easily checked by doing the temperature expansion of the
thermodynamic
potential, which we will do by taking the exact spectrum (\ref{miesp}).

The divergent behavior of the specific heat is related
to the fact
that at the critical temperature  a macroscopic number of
particles suddenly falls
to the ground state. Due to the interaction term, the ground state 
energy is different from zero,

\begin{equation}
E_o=\frac{2\pi \hbar ^2an_o^2}{mV}\left\{ 1+\frac{16}{15}\sqrt{\frac{6a^3
n_o}{\pi V}}\left[ 7E\left( \frac{\pi}{2},\sqrt{\frac{2}{3}}\right) 
-2F\left( \frac{\pi}{2},\sqrt{\frac{2}{3}}\right) \right] \right\} ,
\end{equation}

\noindent  where $E,$ $F$ are the usual elliptic integrals.
Due to interactions there is also a macroscopic number of particles in
states very close to the condensate.

\section{Quasi-particle thermodynamic potential}

The
thermodynamic potential of the quasi-particles whose spectrum was obtained
in the last section
is $\Omega =\frac V{\beta \hbar ^3}\int \frac{d^3{\bf p}}{
(2\pi )^3}\ln [1-\exp (-\beta \varepsilon ({\bf p}))]$
and $\beta =\frac{1}{kT}$. We have taken the chemical potential
$\mu =0$. We
are going to do an asymptotic expansion of this potential close to $T_c$
for $T < T_c$, taking into
account that for $T \to T_c$, $n_o \to 0.$

By changing to the variable $x=\sqrt{\frac{\beta}{2m}}p$
it reads

\begin{equation}
\Omega =\frac{2(2m)^{3/2}V}{\beta ^{5/2}\hbar^3}\int\limits_0^\infty
dx x^2 \ln [1- e^{-\sqrt{12M^2+8Mx^2+x^4}}]. ,\label{omeg}
\end{equation}

\noindent where $M=K\beta \ll 1$ and we are going to do our expansion in
terms of it, in the
same way done in the case of the effective potential in the abelian
Higgs model \cite{Dol}.

Thus,

\begin{equation}
_{}\Omega =\Omega (M=0)+\left. \frac{\partial \Omega }{\partial M}\right |
_{M=0}M+R(M),
\end{equation}

\noindent where we stop our expansion in the first-order term and $R(M)$ is
certain function of $M$ that we are going to find out as

\begin{equation}
R(M)=\Omega (M)-\Omega (M=0)-\left. \frac{\partial \Omega }{\partial M}=
\right| _{M=0}\,M ,
\end{equation}

\noindent if the first derivative of this expression is calculated we
obtain

\begin{equation}
\frac{\partial R}{\partial M}=\frac{\partial \Omega }{\partial M}-\left.=
\frac{\partial \Omega }{\partial M}\right |_{M=0},  \label{R}
\end{equation}

\noindent and this is the expression to be used for computing $R(M).$

After some calculations that are shown in the appendix the thermodynamic
potential is obtained as

\begin{eqnarray}
\Omega (M) &=&\frac{(2m)^{3/2}}{^{}\hbar ^3(2\pi )^2}\left\{ \frac{-2}
{3}\Gamma (\frac{5}{2})\zeta (\frac{5}{2})k_{}^{5/2}T^{5/2}+4\Gamma (\frac
{3}{2})\zeta
(\frac{3}{2})k_{}^{3/2}T^{3/2}\right. \\
&&\ \ \left. -\alpha \frac{8\pi }{3}kTK^{3/2}\right.  \nonumber
\end{eqnarray}
\begin{eqnarray*}
&&\ \ \left. -\frac{ 8}{5}\sqrt{\frac{ 2}{3}}K^{5/2}[7E(\frac \pi
2,\sqrt{\frac{2}{3}}
)-2F(\frac{\pi}{ 2},\sqrt{\frac{ 2}{3}})]\right. \\
&&\left. +\zeta (\frac{ 3}{2})k_{}^{1/2}T^{1/2}\frac{O(K^3)}{12\sqrt{\pi
^3}}%
\right\} +const. \\ \label{tr}
&&
\end{eqnarray*}
Here $\alpha = (\frac{5}{4} + \frac{\sqrt{3}}{2})(\sqrt{6} - \sqrt{2})$.

The most interesting term for us is the third one, containing the infrared
contribution. Such behavior can be obtained also by cutting the interval
of integration in (\ref{omeg}) in the form $\int_0^{\infty} =
\int_0^{\eta} + \int_{\eta}^{\infty}$ by taking some small $\eta \gg M$.
Then in the
first integral we  can expand the exponential in the
denominator, leading to a linear in $T$ term, which contains the
third term of (\ref{tr}).

\section{Internal energy and specific heat}

As $K\to 0$ near $T_c$, in what follows we are
taking into account only the terms up to
$K^{3/2}$. Then for $U=\Omega -T\frac{\partial \Omega }{\partial T}$ we
obtain

\begin{eqnarray}
U &=&\frac{(2m)^{3/2}}{^{}\hbar ^3(2\pi )^2}\left\{ \Gamma (\frac{5}{2})\zeta
(\frac{5}{2})k_{}^{5/2}T^{5/2}-2\Gamma (\frac{3}{2})\zeta (\frac
{3}{2})k_{}^{3/2}T^{3/2}K\right. \\
&&\left. -4\Gamma (\frac{3}{2})\zeta (\frac{3}{2})k_{}^{3/2}T^{5/2}
\frac{\partial
K}{\partial T}+\alpha 4\pi kT^2K^{1/2}\frac{\partial K}{\partial T}\right\},
\nonumber
\end{eqnarray}

\noindent and for the specific heat $c_v=\frac{\partial U}{\partial T}$ 
the expression

\begin{equation}
c_v=\frac{(2m)^{3/2}}{^{}\hbar ^3(2\pi )^2}\left\{ \frac{5}{2}\Gamma (\frac
{5}{2})\zeta (\frac{5}{2})k_{}^{5/2}T^{3/2}-3\Gamma (\frac 32)\zeta (\frac
{3}{2})k_{}^{3/2}T^{1/2}K\right.  \label{cv}
\end{equation}

\begin{eqnarray*}
&&\left. +4kT[2\alpha \pi K^{1/2}-3\Gamma (\frac{3}{2})\zeta (\frac
{3}{2})k_{}^{1/2}T^{1/2}]\frac{\partial K}{\partial T}\right. \\
&&\left. -4kT^2[\Gamma (\frac{3}{2})\zeta (\frac{3}{2})k^{1/2}T^{1/2}-
\alpha \pi K^{1/2}]\frac{\partial^2K}{\partial T^2}\right.
\end{eqnarray*}

\[
\left. +\alpha 2\pi kT^2\left( \frac{\partial K}{\partial T}\right)
^2K^{-1/2}\right\} , \label{29}
\]

\noindent and it is the last term in (\ref{29}) the one giving the
divergent
behavior of $c_v$ for $T \to T_{c-}$. The first term
gives exactly the contribution of the ideal boson gas.

The contribution of the last term can be expressed in terms of
$n_o$ as

\begin{eqnarray}
\triangle c_v &=&\frac{(2m)^{3/2}}{^{}\hbar^3(2\pi )^2}\left\{ \alpha 2\pi
kT^2\left( \frac{2\pi a\hbar ^2}{mV}\right)^{3/2}
\left(\frac{\partial n_o}{\partial T}\right) ^2n_o^{-1/2}\right\}, \\
&&  \nonumber
\end{eqnarray}

By taking
$n_o=N\left[ 1-\left( \frac{ T}{T_c}\right)^\gamma \right] $, ($\gamma >
0$,
for the ideal
gas $\gamma =\frac{2}{3}$), one obtains that for $T \to T_{c-}$,
$\triangle c_v \sim \frac{N \gamma^2 T^{2\gamma}}{T^{2\gamma -2_c}
\sqrt{1-(T/T_c)^{\gamma}}}$.

The behavior of $c_v$ for $T > T_{c}$ would require a separate
investigation. For exactly $T = T_c$, as argued in
a previous
paper \cite{Perez}, for the ideal gas not in
the thermodynamic limit, the number of
particles in
the condensate is not zero, and for $T \simeq T_+$ we may expect
that the densities in states close to the ground state are large, but
rapidly decreasing with increasing $T$. In that region, the chemical
potential cannot be taken as zero. Thus, we expect that a mechanism
similar to the one described for $T < T_c$ may perhaps leads to a
divergent behavior in that region for $T \to T_c$. We shall take, however,
the free gas behavior in our present approximation.
We will write also

\begin{equation}
n_o = Nf(T).
\end{equation}

The specific form of $f(T)$ can be taken as the ideal gas one since
usually the quantity $\alpha_{\bf p} \ll 1$ and thus
the average density of quasiparticles $n_{\bf p}$ is of the same order
of the
particles not in the ground state $N_{\bf p}$ (since $N_{\bf p} =
n_{\bf p}(1 + \alpha^2_{\bf p}/1 - \alpha^2_{\bf p}) + \alpha^2_{\bf p}/1
- \alpha^2_{\bf p}$), \cite{Landau},
and these are proportional to $T^{3/2}$.
We take, thus,

\begin{equation}
f(T)=\left[ 1 - \left( \frac{T}{T_c}\right)^{3/2}\right] ,
\end{equation}

\noindent  along with helium parameters, the divergent curve $
c_v(T)$ for $T=T_c$  shown in figure 1 is obtained. In this case
the scattering length has been chosen as the radius of a hard-sphere as $
a=2.14\AA $ (Kalos,Levesque, and Verlet)\cite{Kal}.
We want to remark that such a divergent behavior cannot be obtained by
using
the usual Bogoliubov's spectrum \cite{Huang}.

\section{Conclusions}

From our results we conclude that some quasi-particle boson spectra having 
nonvanishing temperature-dependent infrared limit, for some specific 
dependences 
of this limit on temperature, we get a $\lambda$-type behavior for $c_v$.
For the case of the Bogoliubov's model of the weakly interacting gas
it is seen that if one does not take the approximation
$n_o\simeq N $, but use instead $n_0 (T)$, one  obtains a quasiparticle
spectrum which is nonvanishing in the long wavelength limit, $\lim_{
{\bf p}\to 0}\varepsilon ({\bf p})=\sqrt{12} K$
where $K$ is proportional to the condensate density
$n_o$. The
nonvanishing value of $K$ is  a manifestation of the symmetry breaking,
and
for $n_o = 0$ the spectrum is reduced to the free particle one,
where $\lim_{^{{\bf p}\rightarrow 0}}\varepsilon ({\bf p})=0$. The
infrared contribution to the internal energy contains a term
proportional to $n_0^{1/2}$. For a dependence of $n_0$ on temperature 
of form $n_0 = N [1 - (T/T_c)^{\gamma}]$, $\gamma > 0$, a
$\lambda$-type divergent
behavior of the specific heat $c_v$ is obtained as $T \to T_c$.

\section{Acknowledgments}
One of the authors (H. P. R.) thanks Professor M. Virasoro,
IAEA and UNESCO for hospitality at the International Centre of Theoretical
Physics.
He also would like to thank his former student R.
Torres Rivero for having called his attention about the possible relation
of
the
$\lambda$ type spectrum with some temperature-dependent mass model. Both
authors
thank A. Amezaga, A. Cabo, S. Fantoni, F. Hussain, C. Montonen, K. Narain
A. Polls and R.
Sorkin for
valuable comments and suggestions.

\section{Appendix}
\subsection{Lowest and first order terms of $\Omega $}
If $\Omega $ is
evaluated for $M=0$ the expression for the thermodynamic
potential is

\begin{equation}
\Omega (M=0)=\frac{-2(2m)^{3/2}V}{3\beta^{5/2}\hbar^3(2\pi )^2}\Gamma
(\frac{5}{2})\zeta (\frac{5}{2}) , \label{P1}
\end{equation}

Then

\[
\frac{\partial \Omega }{\partial M}=\frac{(2m)^{3/2}V}{\beta ^{5/2}\hbar
^3(2\pi )^2}\int\limits_0^\infty \frac{dx
x^2[24M+8x^2]}{\sqrt{12M^2+8Mx^2+x^4}
[\exp (\sqrt{12M^2+8Mx^2+x^4})-1]} ,
\]

from which

\begin{equation}
\left. \frac{\partial \Omega }{\partial M}\right| _{M=0} = \frac{4(2m)^{3
/2}V}{
\beta^{5/2}\hbar^3(2\pi )^2}\Gamma (\frac{3}{2})\zeta (\frac{3}{2}). 
\label{P2}
\end{equation}

\subsection{The $R(M)$ function}

If we substitute in (\ref{R}) the expressions (\ref{P1}) , (\ref{P2}) and
the Matsubara sum

\begin{equation}
\frac{1}{\varepsilon [\exp (\varepsilon )-1]}=\sum\limits_{-\infty
}^\infty
\frac{1}{\varepsilon^2 + 4\pi n^2}-\frac{1}{2\varepsilon },
\end{equation}

\noindent we obtain

\begin{eqnarray}
\frac{\partial R}{\partial M} &=&\frac{8(2m)^{3/2}V}{\beta^{5/2}\hbar^3
(2\pi )^2}\left\{\left[ \int\limits_0^\infty \frac{dx
x^2[3M+x^2]}{12M^2+8Mx^2+x^4}-\int\limits_0^\infty \frac{dx x^4}{x^4}\right
]
\right.  \nonumber   \\
&&\left. +2\sum\limits_{n=1}^\infty \left[ \int\limits_0^\infty \frac{dx
x^2[3M+x^2]}{4\pi n^2+12M^2+8Mx^2+x^4}-\int\limits_0^\infty \frac{dx
x^4}{4\pi n^2+x^4}\right] \right.   \\
&&\left. -\frac 12\left[ \int\limits_0^\infty \frac{dx x^2[3M+x^2]}{
\sqrt{12M^2+8Mx^2+x^4}}-\int\limits_0^\infty \frac{dx x^4}{x^2} \right]
\right\} . \nonumber \\
&&  \nonumber
\end{eqnarray}

Each one of the terms between brackets are calculated separately, thus the
first one is

\begin{eqnarray}
\lbrack ...]_0 &=&\frac{-\alpha \pi }2M^{1/2}, \\
&&  \nonumber
\end{eqnarray}

\noindent where $\alpha
=(\frac{5}{4} + \frac{\sqrt{3}}{2})(\sqrt{6} - \sqrt{2})$.

In the second term as $M$ is very small in our range of temperatures
we can take
the limiting case of $n$ large and expand it in a power series of
$\frac
M{\pi n}$ so

\begin{eqnarray}
\lbrack ...]_1 &=&\frac{-3M}4\frac{\sqrt{\pi }}{\sqrt{n}}+\frac 14\frac{M
^2}{(\pi n)^{3/2}}+.... \\
&&  \nonumber
\end{eqnarray}

The last one, after integrating is,

\begin{equation}
\lbrack ...]_2=-ML+\sqrt{\frac 23}M^{3/2}\left\{ 7E(\frac \pi 2,\sqrt{\frac
23})-2F(\frac \pi 2,\sqrt{\frac 23})\right\},
\end{equation}

\noindent where the divergent integrals have been been regularized by
introducing a cut-off $L$ large enough.

The first term of $[...]_1$ when the sum over $n$ is done is a divergent
series. Thus we take an upper cut-off term at $n=N(L)$ such as
to cancel both divergent terms. At the end we get

\begin{eqnarray}
\frac{\partial R}{\partial M} &=&\frac{8(2m)^{3/2}V}{\beta ^{5/2}\hbar
^3(2\pi )^2}\left\{ -\alpha \frac \pi 2M^{1/2}+2\sum\limits_{n=1}^N\frac{-3M
\sqrt{\pi }}{4\sqrt{n}}+\frac 14\frac{O(M^2)}{\sqrt{\pi ^3}}\zeta (\frac
{3}{2})\right. \\
&&\ \left. -\frac 12[-ML+\sqrt{\frac 23}M^{3/2}(7E(\frac \pi 2,\sqrt{\frac
{2}{3}})-2F(\frac \pi 2,\sqrt{\frac 23}))]\right\} , \nonumber
\end{eqnarray}

then the sum can be approximated by

\begin{equation}
\sum\limits_{n=1}^N\frac 1{\sqrt{n}}\approx \int\limits_0^N\frac{dn}
{\sqrt{n}}=2N^{1/2},
\end{equation}

\noindent in this way by
taking $L=\frac{6}{\sqrt{\pi }}N^{1/2}$, the otherwise divergent terms cancel each other
and after the integration in $M$ has
been performed the value of the $R(M)$ function is

\begin{eqnarray}
R(M) &=&\frac{8(2m)^{3/2}}{\beta ^{5/2}\hbar ^3(2\pi )^2}\left\{ -\alpha
\frac \pi 3M^{3/2}\right.  \nonumber \\
&&\left. -\frac 15\sqrt{\frac 23}M^{5/2}[7E(\frac \pi 2,\sqrt{\frac 23}
)-2F(\frac \pi 2,\sqrt{\frac 23})]\right. \\
&&\ \left. +\zeta (\frac 32)\frac{O(M^3)}{12\sqrt{\pi ^3}}+const\right\}.
\nonumber
\end{eqnarray}

\noindent where $const$ does not depend on $M$.

\newpage\

\section{Figure Caption}

Figure 1: ($c_{v,}T)-$curve obtained from our model, where
the divergent behavior of the specific heat is shown.
For $T > T_c$
the ideal gas curve has been
taken.

\newpage

% GNUPLOT: LaTeX picture
\setlength{\unitlength}{0.240900pt} \ifx\plotpoint\undefined%
\newsavebox{\plotpoint}\fi
\sbox{\plotpoint}{\rule[-0.500pt]{1.000pt}{1.000pt}}%
\begin{picture}(1500,900)(0,0)
\font\gnuplot=cmr10 at 10pt
\gnuplot
\sbox{\plotpoint}{\rule[-0.500pt]{1.000pt}{1.000pt}}%
\put(220.0,113.0){\rule[-0.500pt]{4.818pt}{1.000pt}}
\put(198,113){\makebox(0,0)[r]{33.325}}
\put(1416.0,113.0){\rule[-0.500pt]{4.818pt}{1.000pt}}
\put(220.0,240.0){\rule[-0.500pt]{4.818pt}{1.000pt}}
\put(198,240){\makebox(0,0)[r]{33.33}}
\put(1416.0,240.0){\rule[-0.500pt]{4.818pt}{1.000pt}}
\put(220.0,368.0){\rule[-0.500pt]{4.818pt}{1.000pt}}
\put(198,368){\makebox(0,0)[r]{33.335}}
\put(1416.0,368.0){\rule[-0.500pt]{4.818pt}{1.000pt}}
\put(220.0,495.0){\rule[-0.500pt]{4.818pt}{1.000pt}}
\put(198,495){\makebox(0,0)[r]{33.34}}
\put(1416.0,495.0){\rule[-0.500pt]{4.818pt}{1.000pt}}
\put(220.0,622.0){\rule[-0.500pt]{4.818pt}{1.000pt}}
\put(198,622){\makebox(0,0)[r]{33.345}}
\put(1416.0,622.0){\rule[-0.500pt]{4.818pt}{1.000pt}}
\put(220.0,750.0){\rule[-0.500pt]{4.818pt}{1.000pt}}
\put(198,750){\makebox(0,0)[r]{33.35}}
\put(1416.0,750.0){\rule[-0.500pt]{4.818pt}{1.000pt}}
\put(220.0,877.0){\rule[-0.500pt]{4.818pt}{1.000pt}}
\put(198,877){\makebox(0,0)[r]{33.355}}
\put(1416.0,877.0){\rule[-0.500pt]{4.818pt}{1.000pt}}
\put(220.0,113.0){\rule[-0.500pt]{1.000pt}{4.818pt}}
\put(220,68){\makebox(0,0){0.99994}}
\put(220.0,857.0){\rule[-0.500pt]{1.000pt}{4.818pt}}
\put(423.0,113.0){\rule[-0.500pt]{1.000pt}{4.818pt}}
\put(423,68){\makebox(0,0){0.99996}}
\put(423.0,857.0){\rule[-0.500pt]{1.000pt}{4.818pt}}
\put(625.0,113.0){\rule[-0.500pt]{1.000pt}{4.818pt}}
\put(625,68){\makebox(0,0){0.99998}}
\put(625.0,857.0){\rule[-0.500pt]{1.000pt}{4.818pt}}
\put(828.0,113.0){\rule[-0.500pt]{1.000pt}{4.818pt}}
\put(828,68){\makebox(0,0){1}}
\put(828.0,857.0){\rule[-0.500pt]{1.000pt}{4.818pt}}
\put(1031.0,113.0){\rule[-0.500pt]{1.000pt}{4.818pt}}
\put(1031,68){\makebox(0,0){1.00002}}
\put(1031.0,857.0){\rule[-0.500pt]{1.000pt}{4.818pt}}
\put(1233.0,113.0){\rule[-0.500pt]{1.000pt}{4.818pt}}
\put(1233,68){\makebox(0,0){1.00004}}
\put(1233.0,857.0){\rule[-0.500pt]{1.000pt}{4.818pt}}
\put(1436.0,113.0){\rule[-0.500pt]{1.000pt}{4.818pt}}
\put(1436,68){\makebox(0,0){1.00006}}
\put(1436.0,857.0){\rule[-0.500pt]{1.000pt}{4.818pt}}
\put(220.0,113.0){\rule[-0.500pt]{292.934pt}{1.000pt}}
\put(1436.0,113.0){\rule[-0.500pt]{1.000pt}{184.048pt}}
\put(220.0,877.0){\rule[-0.500pt]{292.934pt}{1.000pt}}
\put(45,495){\makebox(0,0)[r]{$Cv(10^{4} J/K)$}}
\put(828,23){\makebox(0,0){$T/T_c$}}
\put(220.0,113.0){\rule[-0.500pt]{1.000pt}{184.048pt}}
\put(1306,812){\makebox(0,0)[r]{$\lambda-point$}}
\put(1328.0,812.0){\rule[-0.500pt]{15.899pt}{1.000pt}}
\put(321,214){\usebox{\plotpoint}}
\put(321,213.42){\rule{5.541pt}{1.000pt}}
\multiput(321.00,211.92)(11.500,3.000){2}{\rule{2.770pt}{1.000pt}}
\put(344,216.42){\rule{4.577pt}{1.000pt}}
\multiput(344.00,214.92)(9.500,3.000){2}{\rule{2.289pt}{1.000pt}}
\put(363,219.42){\rule{5.059pt}{1.000pt}}
\multiput(363.00,217.92)(10.500,3.000){2}{\rule{2.529pt}{1.000pt}}
\put(384,222.42){\rule{5.300pt}{1.000pt}}
\multiput(384.00,220.92)(11.000,3.000){2}{\rule{2.650pt}{1.000pt}}
\put(406,225.42){\rule{5.300pt}{1.000pt}}
\multiput(406.00,223.92)(11.000,3.000){2}{\rule{2.650pt}{1.000pt}}
\put(428,228.92){\rule{4.577pt}{1.000pt}}
\multiput(428.00,226.92)(9.500,4.000){2}{\rule{2.289pt}{1.000pt}}
\put(447,232.42){\rule{5.300pt}{1.000pt}}
\multiput(447.00,230.92)(11.000,3.000){2}{\rule{2.650pt}{1.000pt}}
\put(469,235.42){\rule{5.059pt}{1.000pt}}
\multiput(469.00,233.92)(10.500,3.000){2}{\rule{2.529pt}{1.000pt}}
\put(490,238.42){\rule{5.300pt}{1.000pt}}
\multiput(490.00,236.92)(11.000,3.000){2}{\rule{2.650pt}{1.000pt}}
\put(512,241.92){\rule{5.059pt}{1.000pt}}
\multiput(512.00,239.92)(10.500,4.000){2}{\rule{2.529pt}{1.000pt}}
\put(533,245.42){\rule{4.818pt}{1.000pt}}
\multiput(533.00,243.92)(10.000,3.000){2}{\rule{2.409pt}{1.000pt}}
\put(553,248.92){\rule{5.300pt}{1.000pt}}
\multiput(553.00,246.92)(11.000,4.000){2}{\rule{2.650pt}{1.000pt}}
\put(575,252.92){\rule{5.300pt}{1.000pt}}
\multiput(575.00,250.92)(11.000,4.000){2}{\rule{2.650pt}{1.000pt}}
\put(597,256.92){\rule{5.059pt}{1.000pt}}
\multiput(597.00,254.92)(10.500,4.000){2}{\rule{2.529pt}{1.000pt}}
\put(618,260.42){\rule{4.577pt}{1.000pt}}
\multiput(618.00,258.92)(9.500,3.000){2}{\rule{2.289pt}{1.000pt}}
\multiput(637.00,265.86)(2.867,0.424){2}{\rule{4.850pt}{0.102pt}}
\multiput(637.00,261.92)(12.934,5.000){2}{\rule{2.425pt}{1.000pt}}
\put(660,268.92){\rule{4.577pt}{1.000pt}}
\multiput(660.00,266.92)(9.500,4.000){2}{\rule{2.289pt}{1.000pt}}
\multiput(679.00,274.84)(2.017,0.462){4}{\rule{4.083pt}{0.111pt}}
\multiput(679.00,270.92)(14.525,6.000){2}{\rule{2.042pt}{1.000pt}}
\multiput(702.00,280.86)(2.358,0.424){2}{\rule{4.250pt}{0.102pt}}
\multiput(702.00,276.92)(11.179,5.000){2}{\rule{2.125pt}{1.000pt}}
\multiput(722.00,285.83)(1.295,0.481){8}{\rule{2.875pt}{0.116pt}}
\multiput(722.00,281.92)(15.033,8.000){2}{\rule{1.438pt}{1.000pt}}
\multiput(743.00,293.83)(1.202,0.485){10}{\rule{2.694pt}{0.117pt}}
\multiput(743.00,289.92)(16.408,9.000){2}{\rule{1.347pt}{1.000pt}}
\multiput(765.00,302.84)(0.681,0.462){4}{\rule{1.917pt}{0.111pt}}
\multiput(765.00,298.92)(6.022,6.000){2}{\rule{0.958pt}{1.000pt}}
\multiput(775.00,308.83)(0.608,0.481){8}{\rule{1.625pt}{0.116pt}}
\multiput(775.00,304.92)(7.627,8.000){2}{\rule{0.813pt}{1.000pt}}
\multiput(786.00,316.83)(0.437,0.487){12}{\rule{1.250pt}{0.117pt}}
\multiput(786.00,312.92)(7.406,10.000){2}{\rule{0.625pt}{1.000pt}}
\multiput(797.83,325.00)(0.487,0.703){12}{\rule{0.117pt}{1.750pt}}
\multiput(793.92,325.00)(10.000,11.368){2}{\rule{1.000pt}{0.875pt}}
\multiput(807.86,340.00)(0.424,1.169){2}{\rule{0.102pt}{2.850pt}}
\multiput(803.92,340.00)(5.000,7.085){2}{\rule{1.000pt}{1.425pt}}
\multiput(812.84,353.00)(0.462,1.811){4}{\rule{0.111pt}{3.750pt}}
\multiput(808.92,353.00)(6.000,13.217){2}{\rule{1.000pt}{1.875pt}}
\put(816.42,374){\rule{1.000pt}{4.095pt}}
\multiput(814.92,374.00)(3.000,8.500){2}{\rule{1.000pt}{2.048pt}}
\put(818.92,391){\rule{1.000pt}{6.745pt}}
\multiput(817.92,391.00)(2.000,14.000){2}{\rule{1.000pt}{3.373pt}}
\put(820.42,419){\rule{1.000pt}{5.300pt}}
\multiput(819.92,419.00)(1.000,11.000){2}{\rule{1.000pt}{2.650pt}}
\put(821.92,441){\rule{1.000pt}{8.432pt}}
\multiput(820.92,441.00)(2.000,17.500){2}{\rule{1.000pt}{4.216pt}}
\put(823.42,476){\rule{1.000pt}{6.504pt}}
\multiput(822.92,476.00)(1.000,13.500){2}{\rule{1.000pt}{3.252pt}}
\put(824.42,544){\rule{1.000pt}{17.345pt}}
\multiput(823.92,544.00)(1.000,36.000){2}{\rule{1.000pt}{8.672pt}}
\put(826.0,503.0){\rule[-0.500pt]{1.000pt}{9.877pt}}
\put(825.42,189){\rule{1.000pt}{148.394pt}}
\multiput(824.92,497.00)(1.000,-308.000){2}{\rule{1.000pt}{74.197pt}}
\put(827.0,616.0){\rule[-0.500pt]{1.000pt}{45.530pt}}
\put(836,186.42){\rule{0.723pt}{1.000pt}}
\multiput(836.00,186.92)(1.500,-1.000){2}{\rule{0.361pt}{1.000pt}}
\put(828.0,189.0){\rule[-0.500pt]{1.927pt}{1.000pt}}
\put(850,185.42){\rule{2.168pt}{1.000pt}}
\multiput(850.00,185.92)(4.500,-1.000){2}{\rule{1.084pt}{1.000pt}}
\put(839.0,188.0){\rule[-0.500pt]{2.650pt}{1.000pt}}
\put(869,183.92){\rule{5.300pt}{1.000pt}}
\multiput(869.00,184.92)(11.000,-2.000){2}{\rule{2.650pt}{1.000pt}}
\put(891,182.42){\rule{5.300pt}{1.000pt}}
\multiput(891.00,182.92)(11.000,-1.000){2}{\rule{2.650pt}{1.000pt}}
\put(913,181.42){\rule{4.818pt}{1.000pt}}
\multiput(913.00,181.92)(10.000,-1.000){2}{\rule{2.409pt}{1.000pt}}
\put(933,180.42){\rule{5.059pt}{1.000pt}}
\multiput(933.00,180.92)(10.500,-1.000){2}{\rule{2.529pt}{1.000pt}}
\put(954,178.92){\rule{5.541pt}{1.000pt}}
\multiput(954.00,179.92)(11.500,-2.000){2}{\rule{2.770pt}{1.000pt}}
\put(977,177.42){\rule{4.818pt}{1.000pt}}
\multiput(977.00,177.92)(10.000,-1.000){2}{\rule{2.409pt}{1.000pt}}
\put(997,176.42){\rule{5.300pt}{1.000pt}}
\multiput(997.00,176.92)(11.000,-1.000){2}{\rule{2.650pt}{1.000pt}}
\put(1019,175.42){\rule{4.577pt}{1.000pt}}
\multiput(1019.00,175.92)(9.500,-1.000){2}{\rule{2.289pt}{1.000pt}}
\put(1038,173.92){\rule{5.300pt}{1.000pt}}
\multiput(1038.00,174.92)(11.000,-2.000){2}{\rule{2.650pt}{1.000pt}}
\put(1060,172.42){\rule{4.818pt}{1.000pt}}
\multiput(1060.00,172.92)(10.000,-1.000){2}{\rule{2.409pt}{1.000pt}}
\put(1080,171.42){\rule{5.300pt}{1.000pt}}
\multiput(1080.00,171.92)(11.000,-1.000){2}{\rule{2.650pt}{1.000pt}}
\put(1102,170.42){\rule{5.059pt}{1.000pt}}
\multiput(1102.00,170.92)(10.500,-1.000){2}{\rule{2.529pt}{1.000pt}}
\put(1123,168.92){\rule{5.059pt}{1.000pt}}
\multiput(1123.00,169.92)(10.500,-2.000){2}{\rule{2.529pt}{1.000pt}}
\put(1144,167.42){\rule{5.300pt}{1.000pt}}
\multiput(1144.00,167.92)(11.000,-1.000){2}{\rule{2.650pt}{1.000pt}}
\put(1166,166.42){\rule{5.059pt}{1.000pt}}
\multiput(1166.00,166.92)(10.500,-1.000){2}{\rule{2.529pt}{1.000pt}}
\put(1187,165.42){\rule{5.300pt}{1.000pt}}
\multiput(1187.00,165.92)(11.000,-1.000){2}{\rule{2.650pt}{1.000pt}}
\put(1209,163.92){\rule{4.577pt}{1.000pt}}
\multiput(1209.00,164.92)(9.500,-2.000){2}{\rule{2.289pt}{1.000pt}}
\put(1228,162.42){\rule{5.541pt}{1.000pt}}
\multiput(1228.00,162.92)(11.500,-1.000){2}{\rule{2.770pt}{1.000pt}}
\put(1251,161.42){\rule{4.818pt}{1.000pt}}
\multiput(1251.00,161.92)(10.000,-1.000){2}{\rule{2.409pt}{1.000pt}}
\put(1271,160.42){\rule{5.059pt}{1.000pt}}
\multiput(1271.00,160.92)(10.500,-1.000){2}{\rule{2.529pt}{1.000pt}}
\put(1292,159.42){\rule{4.818pt}{1.000pt}}
\multiput(1292.00,159.92)(10.000,-1.000){2}{\rule{2.409pt}{1.000pt}}
\put(1312,157.92){\rule{5.300pt}{1.000pt}}
\multiput(1312.00,158.92)(11.000,-2.000){2}{\rule{2.650pt}{1.000pt}}
\put(859.0,187.0){\rule[-0.500pt]{2.409pt}{1.000pt}}
\end{picture}

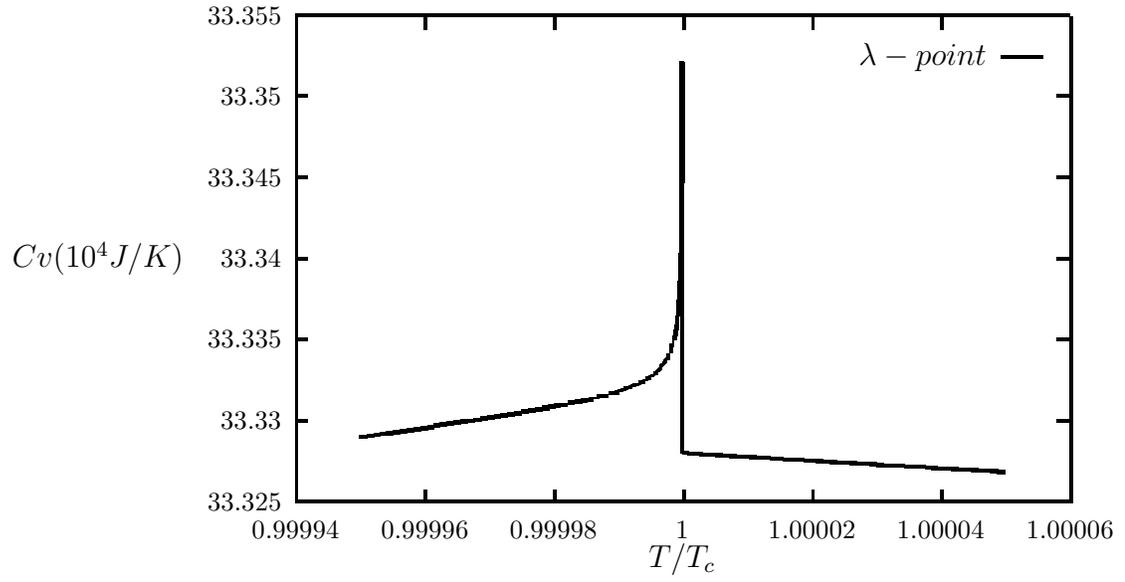
\end{document}